\documentclass[10pt]{iopart}
\usepackage{braket}
\usepackage{graphicx}

\usepackage{soul}
\usepackage{hyperref}
\usepackage{xcolor,colortbl}
\usepackage{todonotes}

\begin{document}
\title[Graphene-based Single-electron Transistors]{Simulating graphene-based single-electron transistor: incoherent current effects due to the presence of electron-electron interaction}

\author{Washington F. dos Santos$^1$,
Felippe Amorim$^{1,2}$ and Alexandre Reily Rocha$^{1,}$\footnote{Author to whom correspondence should be addressed}}

\address{$^1$ Instituto de Física Teórica, S\~ao Paulo State University, S\~ao Paulo, Brasil}
\address{$^2$School of Physics, Trinity College Dublin, Dublin 2, Ireland}
\ead{\mailto{w.santos-junior@unesp.br}, \mailto{alexandre.reily@unesp.br}}

\begin{abstract}
Carbon-based nanostructures have unparalleled electronic properties. At the same time, using an allotrope of carbon as the contacts can yield better device control and reproducibility. 
In this work, we simulate a single-electron transistor composed of a segment of a graphene nanoribbon coupled to carbon nanotubes electrodes. Using the non-equilibrium Green's function formalism we atomistically describe the electronic transport properties of the system including electron-electron interactions. Using this methodology we are able to recover experimentally observed phenomena, such as the Coulomb blockade, as well as the corresponding Coulomb diamonds. Furthermore, we separate the different contributions to transport and show that incoherent effects due to the interaction play a crucial role in the transport properties depending on the region of the stability diagram being considered.  
\end{abstract}

\submitto{Nanotechnology} 
\maketitle
\ioptwocol

\section{Introduction}
Carbon-based structures such as nanotubes \cite{Iijima,Iijima2}, and more recently graphene \cite{Novoselov_2004,Geim2007TheRO,castro_neto_electronic_2009}, have attracted a great deal of attention from the scientific community both from the point of view of basic science \cite{NanotubeTransistors,10-nmGraphene,PhysRevB.97.085413} as well possible applications in photonics \cite{photonics,villegas_optical_2014} and electronics \cite{transpor,TransAlex1,TransAlex2}. In the particular case of graphene, a two-dimensional carbon structure organized in a hexagonal lattice, one is presented with an extremely strong material with unparalleled electrical \cite{castro_neto_electronic_2009,transpor}, and thermal transport properties \cite{thermal}.

One of the drawbacks, however, is the lack of an electronic gap, which hinders several applications in electronics. In that respect, experimental advances have allowed for the fabrication of atomically precise chemically-derived graphene nanoribbons (NR) \cite{Chemically1,Chemically2}. These quasi-one dimensional structures are typically semiconducting with a band gap that can be controlled by ribbon width \cite{widthDependent1}. Another issue that is difficult to overcome is the characterization of the coupling between the device and the metallic contacts. To solve that issue, proposals for systems consisting solely of carbon are becoming more common \cite{Santos,Lin_2016,ZHANG20222497,Zhang2,Zhang3,zhang4,zhang5,Huang}. In that respect, a structure consisting of junctions of carbon nanotubes and a NR has been theoretically proposed by Santos \etal \cite{Santos}. In their work, the authors considered a non-interacting strongly coupled structure, where an unzipped nanotube was used for mimicking the junctions. Recently, Jian Zhang \etal\cite{ZHANG20222497} have built on that idea to construct a multi-gate device architecture using single-wall carbon nanotubes (SWNTs) as electrodes \cite{ZHANG20222497}. The authors synthesized long armchair semiconducting NRs of lateral width $N_{a} = 9$ and deposited them on SWNTs of approximately 1 nm in diameter. Using their approach the nanotubes exhibit self-alignment, with separations of
roughly 15 to 25 nm and a diameter of about
1 nm. Via this technique, they were able to achieve a system with dimensions smaller than those attainable by current synthesis and lithography methods. This SWNT-NR-SWNT setup presents the properties of single-electron transistors at low temperatures. In particular, the authors observed Coulomb blockade and the corresponding Coulomb diamonds \cite{ZHANG20222497}, resulting in a single electron-transistor regime. While the presence of this effect had been previously observed in graphene-based structures \cite{PhysRevLett.108.266601,Stampfer_2008}, there the coupling with the electrode was much better controlled, thus paving the way toward reproducibility. 

In order to fully describe these devices, a proper theoretical description considering the presence of electron-electron interactions is mandatory. This is typically neglected \cite{Santos} in electronic transport simulations of carbon-based systems, as it involves solving a many-body problem. Techniques with different levels of approximation have been developed to tackle this issue \cite{Claveau_2014,Hewson_2001,Bartosch_2009,Haule_2001,Haule_2001,melo_quantitative_2019}, 
focusing on a small number of interacting orbitals \cite{melo_quantitative_2019}. This, in principle, is not the case here, as the quantum-dot-like behavior of the system arises from the weak coupling between nanotube and nanoribbon, and the relevant states can be spatially delocalized over the entire ribbon. 

In this work, we simulated a single electron transistor based on carbon nanotubes as electrodes, coupled to a graphene nanoribbon, which is similar to the experimental setup \cite{ZHANG20222497}. Here, we seek to understand how the electron-electron interaction affect the electronic and transport properties of this device. To this end, we perform calculations using the non-equilibrium Green's function method \cite{smeagol}, describing the interacting states via a multi-orbital Anderson impurity model (AIM) with the non-crossing approximation (NCA) \cite{de2019quantitative} and a generalization to the Meir-Wingreen formula for incoherent transport \cite{Generalized}. We map the many-orbital Hamiltonian into a 2-level impurity model, and via the combination of these methods, we correctly describe transport phenomena for interacting systems, including Coulomb diamonds. We find that we are able to account for the Coulomb diamonds, and most importantly to separate the different contributions of the current to show that the incoherent contributions dominate the transport for a wide range of applied external bias and gate voltages.

\section{Methodology} 

\begin{figure*}[ht]
\centering
\includegraphics[width=1.0 \textwidth]{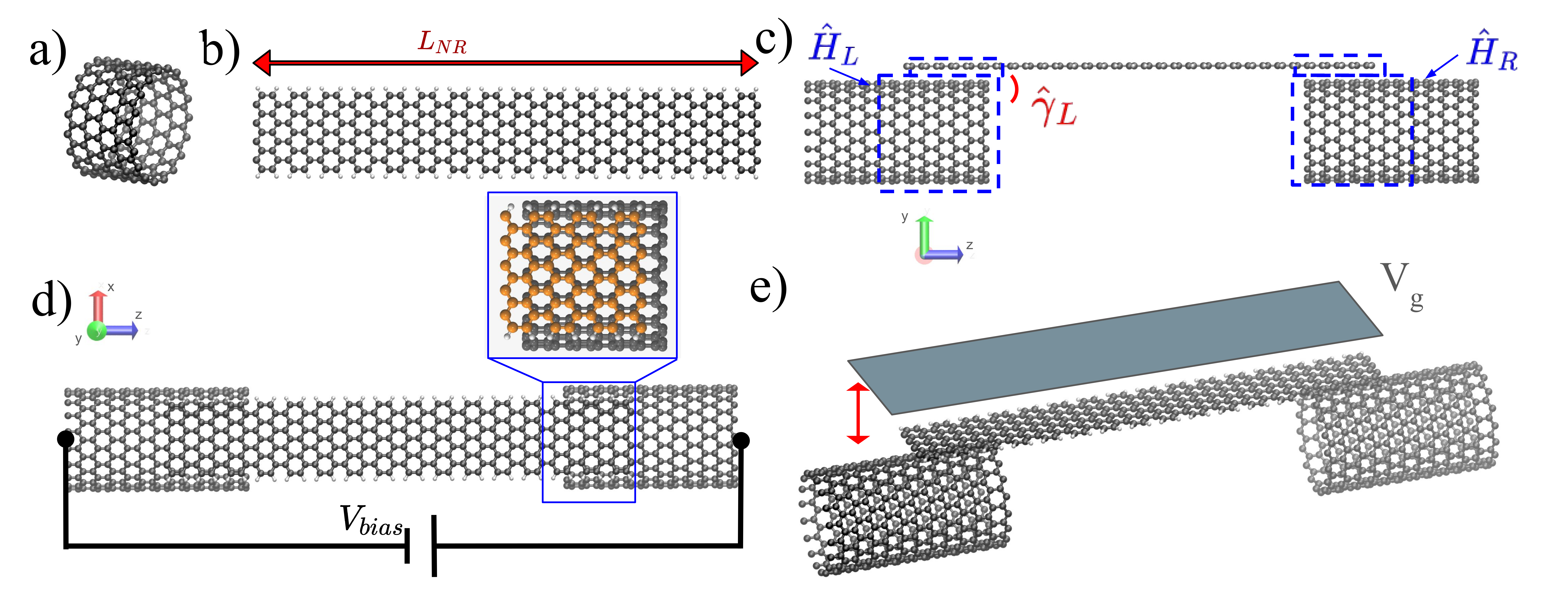}
\caption{Segment of a (18,0) nanotube. b) 9-zigzag graphene nanoribbon of length 16 hexagons.  c) Side and d) top views of the SWNT-NR-SWNT device. Three layers of the NR on each side of the NR are coupled to the nanotube with AA stacking (zoomed-in view). The dashed red line shows one unit cell of the electrode. e) Representation of a gate potential applied to the nanoribbon to shift the on-site energy levels.}
\label{estr}
\end{figure*}

The setup used throughout this work is presented in Fig.~\ref{estr}(c-d), and follows the arrangement observed in experiments \cite{ZHANG20222497}. We place two semi-infinite (18,0) zig-zag SWNTs (metallic) separated by a distance of approximately 20 \AA. These make up the left and right electrodes. We then take a finite segment of a 9-armchair graphene nanoribbon (A-NR),  consisting of $L_{\rm NR} = 16$ hexagons. This is the same width used in the experimental work of Zhang  {\it et al.}, however, the experimental length is approximately 30 nm \cite{ZHANG20222497}. We make the edges of the NR partially overlap with each of the electrodes (three unit cells on either side of the ribbon), and the systems interact via weak van der Waals interactions. The structures considered here are commensurate. We have tested different types of stacking and overlap between nanotube and ribbon (see supplementary material), with similar results for all of them, thus here we considered the simpler AA-stacking.

The system is described using a tight binding Hamiltonian with a single carbon $p_{z}$ orbital per site. We divide our system into three regions as first described by Caroli \etal\cite{Caroli_1971,Caroli_II}, namely left and right semi-infinite electrodes and a scattering region. The latter consists of the NR and the finite nanotube segments that overlap with it.

The Hamiltonian of the scattering region $\hat{H}_{S}$ can be written as, 
\begin{equation}
\hat{H}_{S} = \hat{H}_{L} + \hat{H}_{\rm NR} + \hat{H}_{R} + \hat{\gamma}_{L} + \hat{\gamma}_{R},
\end{equation}
where $H_{L}$ and $H_{R}$, represent the structure of a finite segment of the SWNTs with four unit-cell layers, $H_{\rm NR}$ corresponds to the NR, and $\hat{\gamma}_{L / R}$ is the van der Waals coupling matrix between the corresponding nanotube segment and the NR.  For simplicity, we will assume only first nearest neighbor coupling, and given the AA-stacking, the matrix elements are equal between sites of the NR and SWNT, which are on top of each other (Fig.~\ref{estr}(d)), and zero otherwise.

The tight-binding parameters of the system were chosen according to the literature \cite{Paramtb}. Thus, we take the onsite energy $\epsilon_{0} = 0$ eV, the nearest neighbor coupling $t = 2.7$ eV, and the coupling terms between orbitals at the edges of the NR, are slightly altered to $t_b = t + 0.12t$ to account for saturation of the bonds at the edges. The effective van der Waals coupling in the central region of the NR was taken to be $\gamma_{L / R} = 0.15$ eV as the average value. This is smaller, but comparable to the one in a graphene bilayer ($\approx 0.3$ eV \cite{vdWc}). The different value was taken in order to account for the decrease in coupling at the edge of the ribbon due to the curvature of the nanotube. We can also apply a potential bias $V_{bias}$ between the electrodes, as can be seen in Fig.~\ref{estr}(d) and a gate voltage $V_{g}$ shifting the NR on-site levels (see Fig.~\ref{estr}(e)) 

\begin{figure}[!ht]
\centering
\includegraphics[scale=0.325]{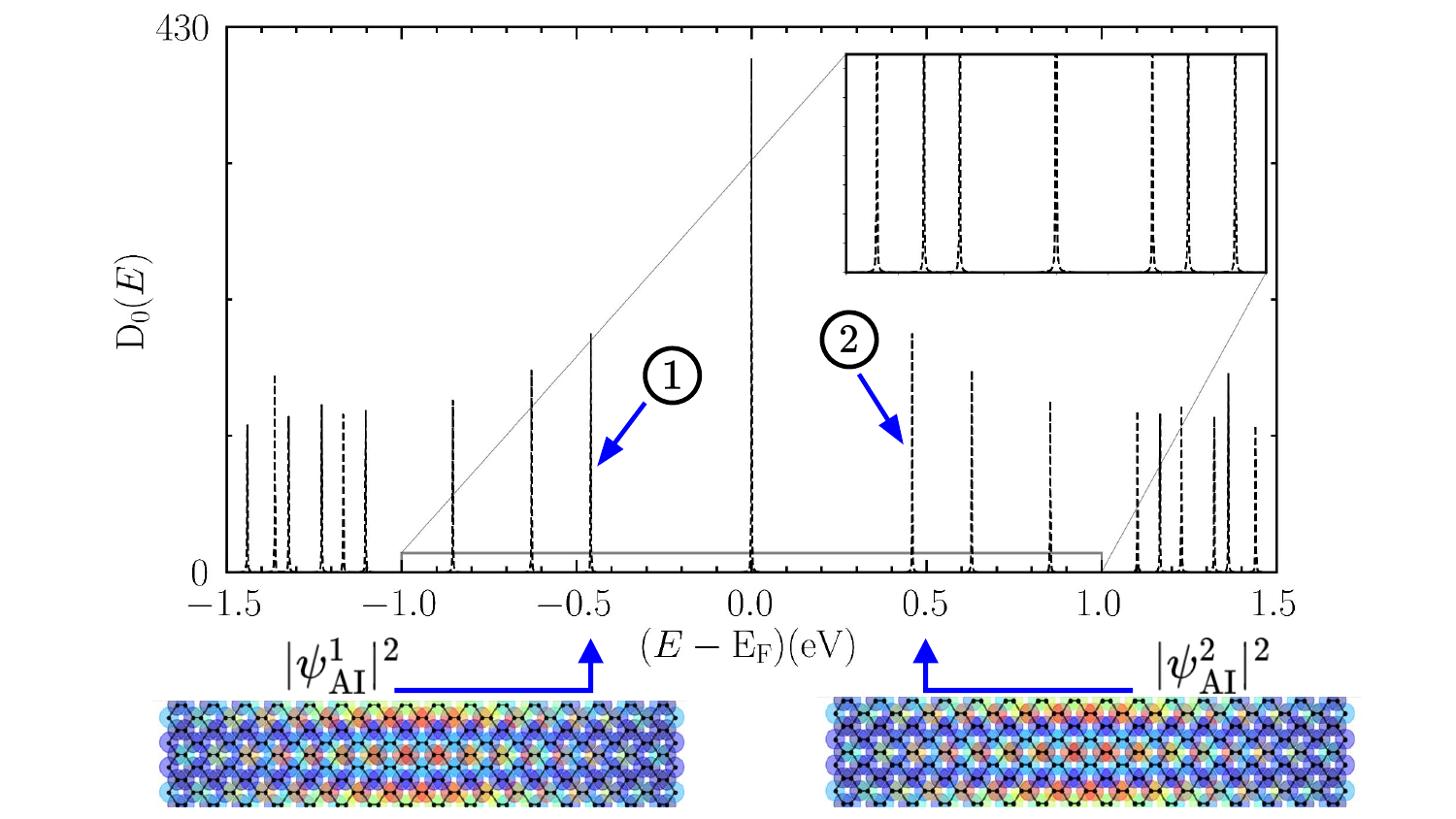}
\caption{\textbf{Density of States for uncoupled NR}: density of States for uncoupled
NR (black dashed) and the absolute square of the wave function for the first two molecular states symmetric with respect to the Fermi-level: $\epsilon^{1}_{\rm AI} = -0.45\,$eV, $\epsilon^{2}_{\rm AI} = 0.45\,$eV.}
\label{stats}
\end{figure}

Starting off from the non-interacting system, in equilibrium, the retarded Green's function \cite{Dattamesoscopic,transiesta,Rocha2006}
\begin{equation}
    G_{0} = \lim_{\eta \to 0^+} [(E+i\eta)I- H_{S} - \Sigma_{L}-\Sigma_{R}]^{-1},
\end{equation}
is the key quantity, where  $\Sigma_{L}$ ($\Sigma_{R}$) is the so-called self-energy for the left (right) electrode. Given the crystal structure of the electrodes, and using Dyson's equation, the self-energies can be obtained via a recursive method \cite{pedago}. 

From this, the density of states (spectral function) projected onto atomic orbitals of the NR can be obtained
\begin{equation}\label{D}
D_{0}(E) = -\frac{1}{\pi} \Tr \big[ Im \{G_{0}\}_{\rm NR} \big],
\end{equation}
and the non-interacting current is given by the Landauer-Büttiker formula,
\begin{equation}
    I_{0} = \frac{2e}{h}\int \Tr[\Gamma_{R}G^{\dagger}_{0}\Gamma_{L}G_{0}](f_{R}-f_{L}){d E},
\end{equation}
where the $\Gamma_{L/R}$ matrices are defined by
\begin{equation}
    \Gamma_{L/R} = i(\Sigma_{L/R} - \Sigma^{\dagger}_{L/R}),
\end{equation}
$f_{L/R}$ is the Fermi distribution with chemical potential $\mu_{L/R} = \mu_{0} \pm V_{bias}$, $\mu_{0}$ a fixed chemical potential, and the term between square brackets is the transmission, which includes only coherent processes. 

In the case of an isolated segment of the NR, one obtains quantized molecular states, as we observe in Fig.~\ref{stats}. Furthermore, the molecular orbitals close to the Fermi level are delocalized over the entire structure. The coupling between NR and the SWNT is weak, thus the overall effect of connecting the electrode segments is a small broadening of the states. This indicates that it retains most of the molecular character of the uncoupled system. It also means that applying a bias or an external gate can lead to significant changes to its charge state. In that case, as already pointed out, electron-electron interactions can have a significant impact.

In principle, including interactions would require a many-body calculation considering the full Hamiltonian, most notably, the whole scattering region. Here, to tackle this issue, we first define the basis of so-called Anderson impurity molecular orbitals, containing $N_{\rm AI}$ states. These molecular orbitals constitute the interacting subset of the NR molecular states, given by,
\begin{equation}
    U_{\psi} = (\psi^{1}_{\rm AI},\psi^{2}_{\rm AI}, \dots ,\psi^{n}_{\rm AI})^{T}.
\end{equation}

The electron-electron interaction is included by considering a multi-orbital Anderson impurity model for the selected $\psi^{i}_{\rm AI}$, with the non-interacting part given by 
\begin{equation}
   \hat{H}_{\rm NR}\ket{\psi^{i}_{\rm AI}} = \epsilon^{i}_{\rm AI}\ket{\psi^{i}_{\rm AI}}.
\end{equation}
while the electron-electron interaction is modeled by 
\begin{equation}
 \hat{H}^{ee} =  \sum^{N_{\rm AI}}_{i,j,k,l}U_{i,j,k,l} \hat{a}_{i,\sigma}^{\dagger}\hat{a}_{j,\sigma^{\prime}}^{\dagger}\hat{a}_{l,\sigma^{\prime}}\hat{a}_{k,\sigma},
\end{equation} 
where the creation (annihilation) operators $\hat{a}_{i,\sigma}^{\dagger}$($\hat{a}_{i,\sigma}$) act on the i-th AI molecular orbital, and $\sigma$ is the spin projection ($\sigma = \pm$). For simplicity, we take
\begin{equation}
U_{i,j,k,l}=\cases{U&if $l=k=j=i$\\
0&otherwise\\},
\end{equation}
with $U = 0.45$ eV, as an average value based on the experimental range ($0.3$ eV - $0.55$ eV) \cite{Zhang2}. As we are considering that the orbitals are orthogonal to each other, the more general procedure for projection \cite{projec} can be simplified, and we can directly project the interacting Green's function of the scattering region onto the AI states
\begin{equation}
    G_{\rm AI} = \left({\begin{array}{c}
   0_{L} \\
 U_{\psi}  \\
 0_{R}
\end{array}}\right)^{\dagger}\left[G^{-1}_0 -\Sigma_{ee}\right]^{-1}\left({\begin{array}{c}
   0_{L} \\
 U_{\psi}  \\
 0_{R}
\end{array}}\right),
\end{equation}
which includes the electron-electron interaction via the self-energy $\Sigma_{ee}$. At the same time, we define the Green's function for the AI states
\begin{equation}
    G_{\rm AI} = [(E +i\eta)I_{N_{\rm AI}} - \epsilon_{\rm AI} - \overline{\Delta}_{\rm AI} - \overline{\sigma}_{ee}]^{-1},
\end{equation}
with the so-called hybridization function $\overline{\Delta}_{\rm AI}$, and the e-e selfenergy $\overline{\sigma}_{ee}$ on the basic of interacting molecular orbitals. 
 
The coupling via van der Waals interactions between the NR and segments of SWNTs result in weak coupling between the states of the AI and the electrodes. The hybridization function can be used in an impurity solver \cite{melo_quantitative_2019} within the dynamic mean field theory formalism (DMFT) \cite{kotliar_compressibility_2002,kotliar_electronic_2006}. The solver will yield $\overline{\sigma}_{ee}$, and we can then perform an inverse projection to obtain the self-energy in the atomic orbital representation,
\begin{equation}
    \Sigma_{ee} = U_{\psi}\overline{\sigma}_{ee}U^{\dagger}_{\psi}.
\end{equation} 
This defines the set of DMFT equations, which can be iterated self-consistently. 

With the information for the interaction self-energy, we can then calculate the electronic and transport properties for the interacting problem in the low bias limit. Recently we have proposed an {\it ansatz} to generalize the Meir-Wingreen transmission \cite{FMW} for the non-proportional coupling regime - the case here - in such a way that the interacting
current is in a Landauer-Buttiker form \cite{Generalized} 
\begin{equation}\label{uncorrelated_i}
I =\frac{2e}{h}\int dE ~ T_{\rm int}\left(E\right) (f_{R} - f_{L}).     
\end{equation}
In the case of weak coupling with the electrodes, this effective transmission can be approximated by
\begin{eqnarray} \label{uncorrelated_t}
	T_{\rm int} & = & Tr\left[\Gamma_{R}G^{r}\Gamma_{L}G^{a}\right] +  Tr\left[\Gamma_{R}G^{r}\Gamma_{ee}G^{a}\right],   
\end{eqnarray}
with $\Gamma_{ee} = i(\Sigma_{ee} - \Sigma^{\dagger}_{ee})$, and we can identify the effective transmission by a coherent and an incoherent term. Subsequently, we also have the total current $I = I_{\rm{coh}} + I_{\rm{incoh}}$. 

Finally, in order to address the local current we can write the lesser Green's function, given by 
\begin{equation}\label{lesser}
    G^{<} = G^{r}\Sigma^{<}G^{a} = G^{r}\Sigma_{0}^{<}G^{a} + G^{r}\Sigma_{ee}^{<}G^{a},
\end{equation}
with $\Sigma_{0}^{<} = \Gamma_Lf_L + \Gamma_Rf_R$, and the interacting lesser self-energy in the same weak-coupling approximation \cite{Generalized} is Meir-Wingreen for weak coupling, given by
\begin{equation} \label{ua}
[\Sigma_{ee}]_{ij}^{<}=i\frac{[G(\Gamma_{R}f_R+\Gamma_{L}f_L)G^{\dagger}]_{ij}}{[G(\Gamma_{R}+\Gamma_{L})G^{\dagger}]_{ij}}[\Gamma_{ee}]_{ij}.
\end{equation}
From this, we can also obtain the generalization of the local current density 
\begin{equation}\label{locadenscurrent} 
    i_{ij}(E) = \frac{2e}{h} Im \{ H_{ij}G_{ij}^{<}\}. 
\end{equation} 
Again, from equation \ref{lesser}, we can separate the local current density as a combination of coherent and incoherent terms, $i_{ij}(E) = i^{\rm{coh}}_{ij}(E) + i^{\rm{incoh}}_{ij}(E)$.

\section{Results}

\begin{figure}[!ht]
\centering
\includegraphics[width=8.3cm]{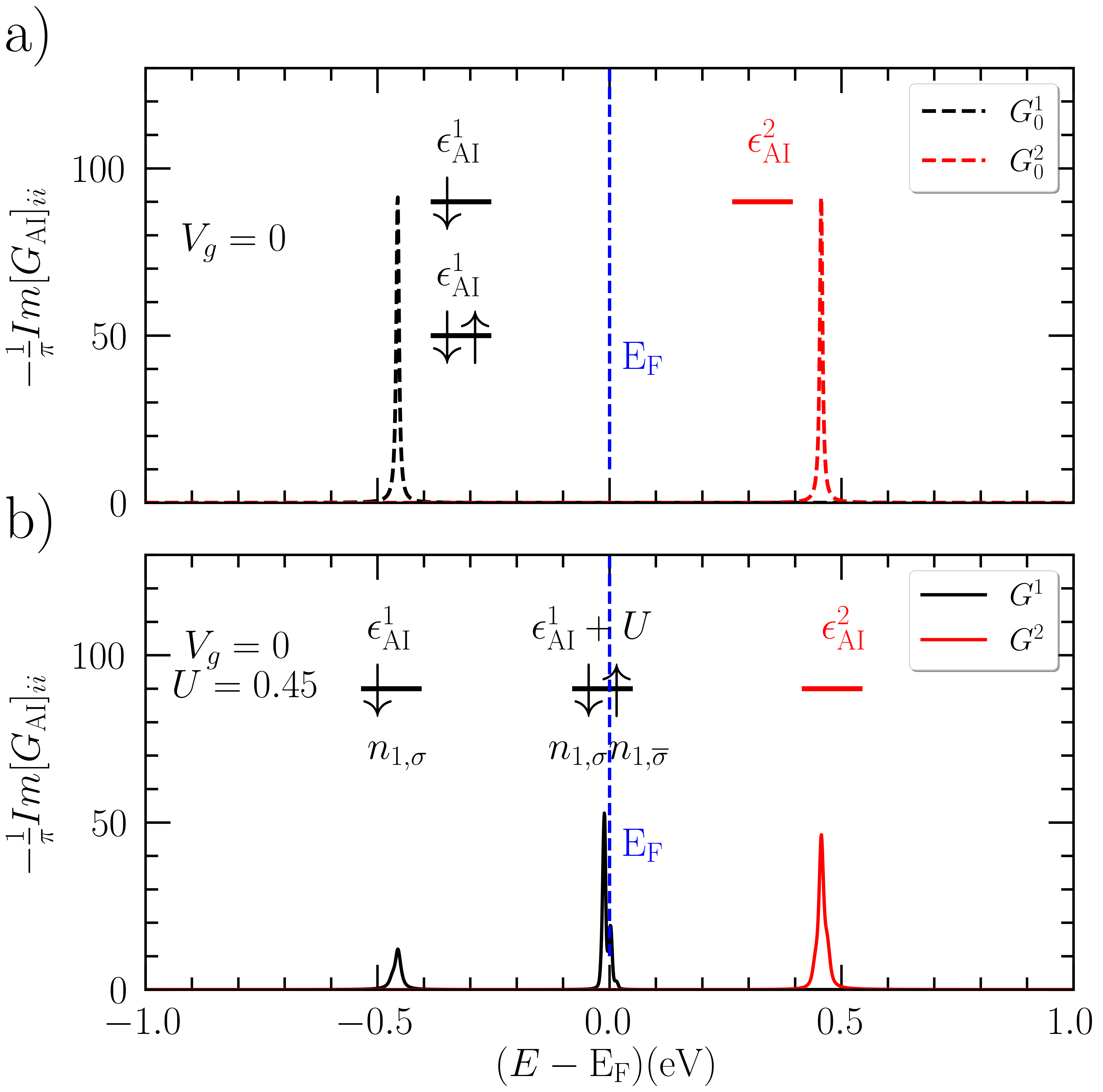}
\caption{\textbf{Projected density of states for Anderson impurities}: density of states for the first two levels close to $\rm{E_{F}}$. In the non-interacting case (a), the doubly-occupied state (black dashed) and zero-occupied state (red dashed). For the interacting ($U=0.45\,$eV) (b) case, the solid line, we have for the single-particle representation zero-occupied state in $E=0.45\,$eV (red solid line), the singly-occupied state in $E=-0.45\,$eV (black solid line) and doubly-occupied state in $E=0$ (black solid line).}
\label{gaiele}
\end{figure}

Following the experimental setup \cite{ZHANG20222497}, we fixed the temperature of the system to $9$K. As already pointed out, given the weak coupling in the junction, we expect the system to retain much of the molecular character. Thus, we select a subset of molecular orbitals, in which we included electron-electron interactions. The choice of interaction states is arbitrary, but we expect those closest to the Fermi level, $\rm{E_{F}}$, to be the most relevant for transport. Here we consider the two states that are symmetrically positioned with respect to Fermi-level ($\epsilon^{1}_{\rm AI}= -0.45\,$eV, $\epsilon^{2}_{\rm AI} = 0.45\,$eV), thus setting $N_{\rm AI} = 2$. We initially also considered the two degenerate states at the Fermi-level, but our calculations indicate that the states don't contribute to the transmission. 

\begin{figure}[!ht]
\centering
\includegraphics[width=8.3cm]{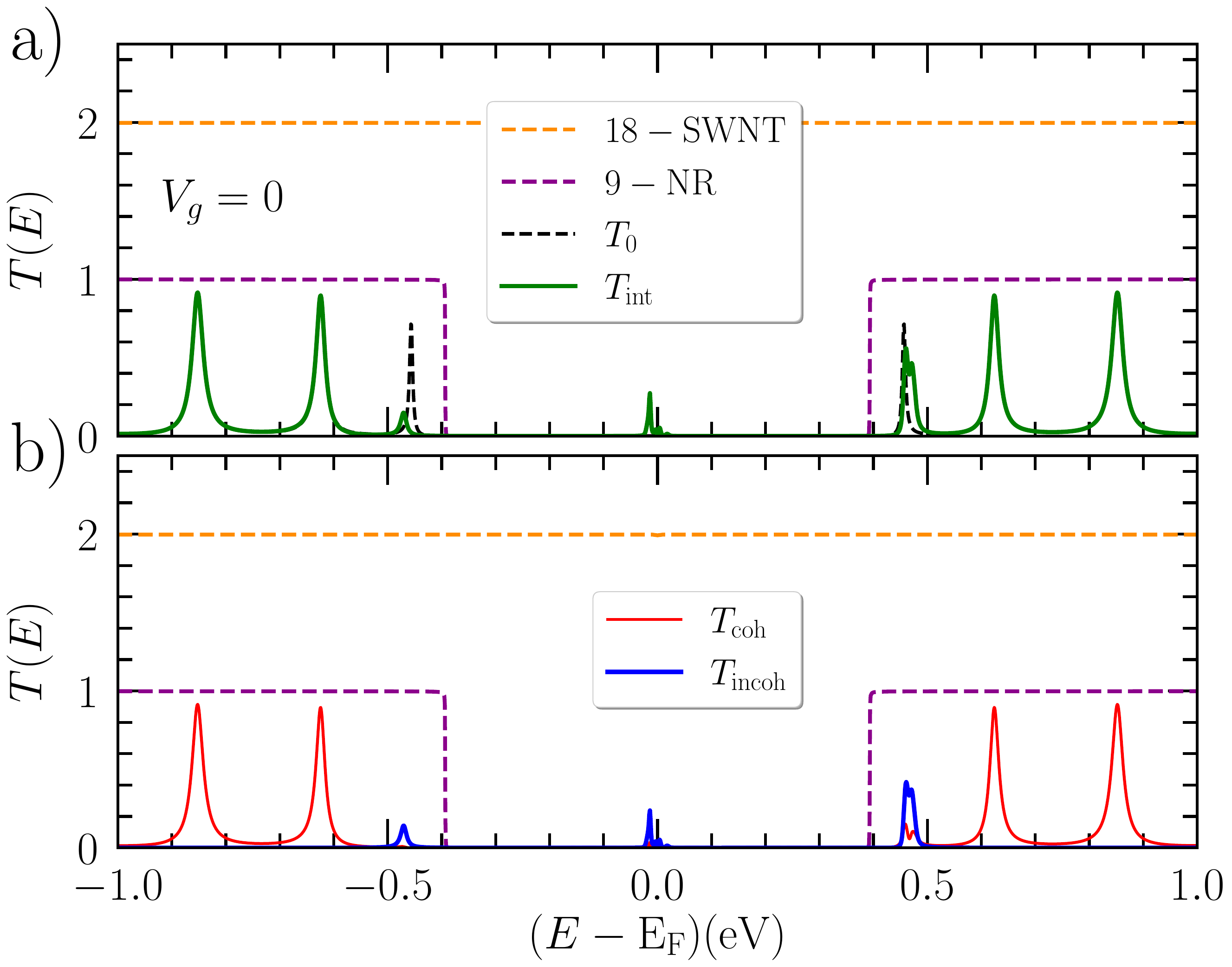}
\caption{\textbf{Transmission as a function of energy}: a) Non-interacting (black solid line) and the interacting (green solid line) transmission b) Coherent (red solid line) and incoherent (blue solid line) transmission terms for interacting SWNT-NR-SWNT with Coulomb potential $U = 0.45$ eV. Transmission for pristine nanotube (orange dashed line) and nanoribbon (purple dashed line) are present as a guide for comparison.}
\label{transele}
\end{figure}

\begin{figure*}[!ht]
\centering
\includegraphics[width=14cm]{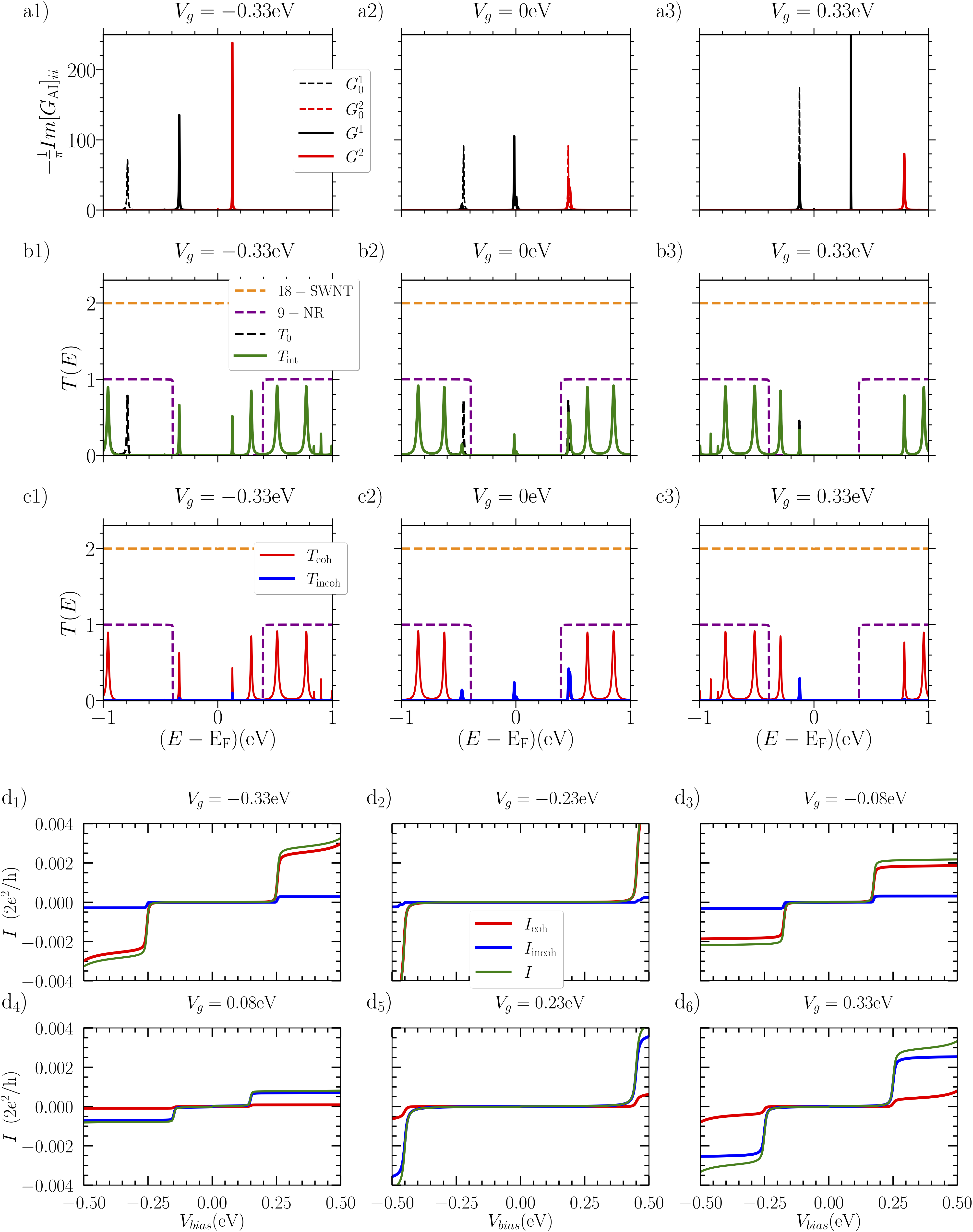}
\caption{\textbf{Density of states, transmission and Current for SWNT-NR-SWNT:} a) Density of states projected on the first two states close to the Fermi-level for increasing gate voltages. For the non-interacting case, the black dashed line refers to $\psi^{1}_{\rm AI}$ having always double-occupation and the red dashed line to $\psi^{2}_{\rm AI}$ with zero-occupation. For the interacting case ($U = 0.45\,$eV), the black solid line refers to $\psi^{1}_{\rm AI}$, and the red solid line refers to $\psi^{2}_{\rm AI}$. b) Zero-bias total transmission as a function of energy for different gate voltages, for non-interacting (black dashed line) and the interacting (green solid line) cases for SWNT-NR-SWNT. c) Coherent (red line) and incoherent (blue line) components of transmission for different gate values for the interacting cases. The Transmission for a pristine nanotube (orange dash line) and a pristine nanoribbon (purple dash line) are present as a guide. d) Comparison between interacting (green line), coherent (red line) and incoherent (blue line) current terms, for $U = 0.45\,$eV as a function of $V_{bias}$ for different gate voltages.}
\label{ALL}
\end{figure*}

Fig.~\ref{gaiele} shows the orbital-projected spectral function for both the interacting and non-interacting cases. We define $G^{i}$ and $G^{i}_{0}$ as the i-th diagonal element of $G_{AI}$ with and without interaction, respectively. In the non-interacting case (Fig.~\ref{gaiele}(a)),  we observe, as expected, that they correspond to weakly coupled single-particle molecular states of the NR with a small broadening. Furthermore, the level just below $\rm{E_{F}}$ remains doubly occupied. When we turn on the interaction (Fig.~\ref{gaiele}(b)) we notice that, as expected \cite{melo_quantitative_2019}, this configuration has one extra peak slightly below $\rm{E_{F}}$. It corresponds to the doubly-occupied level, which is shifted by $U$ and now lies very close to the Fermi level. For this particular choice of $U$, and given the initial position of $\epsilon^{1}_{\rm AI}$, the average occupation on the AI remains approximately 2.

We then calculate the influence of the charged state on the transmission. As shown in Fig.~\ref{transele}(a) the transmission probability close to the Fermi level arises due to the doubly occupied state in the first molecular orbital, which is not present in the non-interacting case. Furthermore, when we look into the different contributions to this peak close to the Fermi level, we notice that it comes mostly from the incoherent term in the transmission. It is important to point out here that, while the coherent term in the interacting case has the same functional form as the Fisher-Lee relation for the non-interacting one, we consider the full interacting Green's function, and therefore many-body effects are also present. This means that, in the regime considered here - above the Kondo temperature - we have to include both coherent and sequential tunneling contributions.

Furthermore, by considering an external gate, it is possible to control the positions of the different levels. This is shown in Fig.~\ref{ALL}.
For the spectral function projected on to the two AI states at negative gate voltages, the behavior of the system can be described by a mean-field approach for the interaction: the completely filled impurity level is shifted upwards by $U$, and this is the only contribution to the transmission coming from that level (see Fig.~\ref{ALL}(b)). For positive values, as the voltage increases, the doubly-occupied level is initially pinned to $\rm{E_{F}}$, but subsequently, it is suppressed in the transmission therefore giving no contribution to the overall transport properties of the system. In the intermediate region, the transmission peak can only be accounted for if the incoherent term is included, this is corroborated by the fact that the largest contribution to the zero-bias transmission comes from the incoherent term, as shown in Fig.~\ref{ALL}(c).

\begin{figure}[!ht]
\centering
\includegraphics[width=8.3cm]{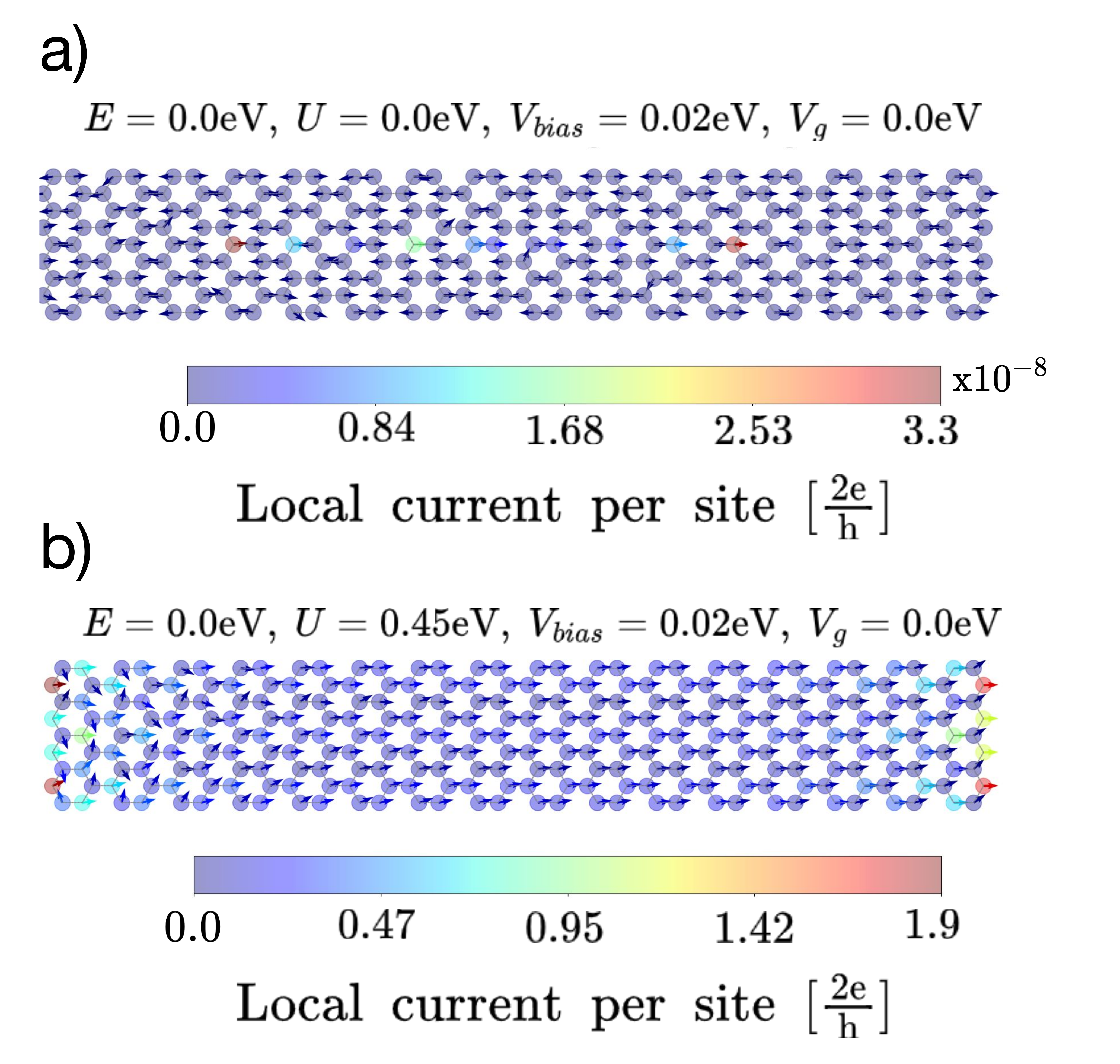}
\caption{\textbf{Local current}: Local current per site $i$, $i_{i}\left(E\right) = \sum_{j}(i_{ij}\left(E\right)\hat{r}_{ij}+i_{j,i}\left(E\right)\hat{r}_{j,i})$ for $E = 0\,$eV, $V_g = 0\,$eV, and $V_b = 0.02\,$eV. a) In the absence of interaction, and b) in the presence of interaction with $U = 0.45\,$eV.}
\label{localcurrent}
\end{figure}

We can also apply an external bias to the device and look away from the strictly zero-bias case. Here, this is done by considering that the interacting self-energies and Green's functions do not change significantly as the bias is applied. Thus, the current is given by integrating equation \ref{uncorrelated_i} within the appropriate bias window, where the local current is given by equation \ref{locadenscurrent}.
We first look at the local current shown in  Fig.~\ref{localcurrent} for $E = 0\,$eV, $V_g = 0\,$eV and $V_{bias} = 0.02\,$eV, we can have a better understanding of the effects of the doubly occupied state on the current density at low bias. In the absence of interaction, we have a current that is zero locally. When we consider electron-electron interactions (Fig.~\ref{localcurrent}(b)), due to the presence of the double-occupied state, the electronic density allows coupling between the electrodes, giving rise, for this small potential difference, to a local current across the quantum dot.

\begin{figure}[!ht]
\centering
\includegraphics[width=8.3cm]{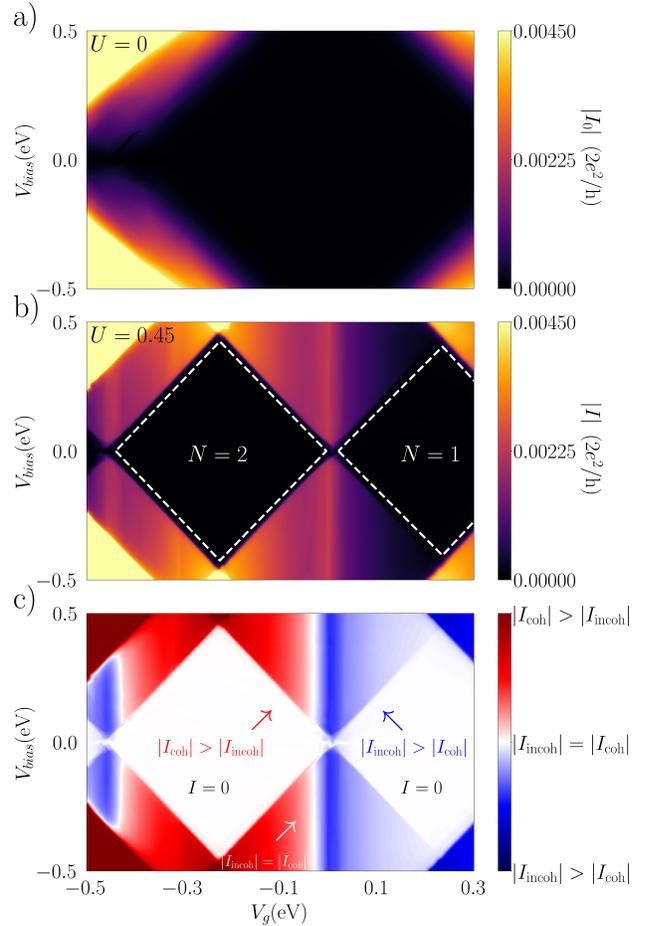}
\caption{\textbf{The stability diagram for current and Difference between current terms, $|I_{\rm{coh}}| - |I_{\rm{incoh}}|$}: a) The stability diagram for current, with $U = 0\,$eV: the absolute of current as a function of gate (x-axis) and bias (y-axis). b) The stability diagram for effective current, with $U = 0.45\,$eV: the absolute of current as a function of gate (x-axis) and bias (y-axis). c) Diagram for difference between absolute of current terms, with $U = 0.45\,$eV as a function of gate (x-axis) and bias (y-axis). In the red region, $|I_{\rm{coh}}|$ is the dominant term, in the blue region, $|I_{\rm{incoh}}|$ is the dominant term and in the white region we have $|I_{\rm{coh}}|$ = $|I_{\rm{incoh}}|$.}
\label{figdiamond}
\end{figure}

Finally, for the total current, we observed in Fig.~\ref{ALL}(d) that there is a competition between the coherent and incoherent terms. For some gate voltages $I_{\rm{coh}}$ is the dominant term, however for others the contribution to the $I_{\rm{incoh}}$ is more intense. These results also corroborate the argument that a mean-field description is valid when the level is doubly occupied ($V_{g}>0$).

By sweeping both the bias and $V_g$, we can obtain the stability diagram, \textit{i.e }the current as a function of bias and gate voltage and compare the non-interacting (Fig. \ref{figdiamond}(a)) and interacting cases (Fig. \ref{figdiamond}(b)). We can then observe that, for $U = 0\,$eV, the current in the central region of the diagram is zero. When we consider $U = 0.45\,$eV, we note the emergence of Coulomb diamonds and a good description of Coulomb blockade for the zero-current regions. This is in accordance with the experimental results \cite{ZHANG20222497} and demonstrates that, for low bias, we are in the single-electron transport regime. Finally, in Fig.~\ref{figdiamond}(c) we present the difference between $\left|I_{\rm{coh}}\right|$ and $|I_{\rm{incoh}}|$. In the region where $\Gamma_{ee} = \rm Im\left[\Sigma_{ee}\right]$ is small, the current is dominated by coherent transport, where $\rm Im\left[\Sigma_{ee}\right] \neq 0$ the incoherent term is leading one. There, one observes the emergence of decoherence phenomena. In fact, the same effect has been noted by Lozano-Negro \etal \cite{Lozano-Negro_2022} in the whole system. In essence, we are able to map the phase space of transport dependent on externally controlled parameters. On a final note on temperature effects, given the energy scales involved, one expects no significant changes in the interacting self-energy up to room temperature (see supplementary information). Nonetheless, it also contributes to the broadening of the Fermi distribution in the calculation of the current. There, it has a large effect, and the sharp changes in current lead to the Coulomb diamonds being washed out, even though the effects due to interaction are still present.

\section{Conclusions}

In this work, we studied the effect of electron-electron interactions in graphene-based devices. At first, we noticed that considering the presence of interaction in molecular states instead of atomic orbitals provides a better physical description, and simplifies the problem. The presence of the electron-electron interaction proved to be an essential rule for the description of this system where Coulomb blockade is present. This can only be done because we accounted for the incoherent transport term. Besides the Coulomb diamond, we also showed the existence of regions where the incoherent process governs the dynamics of electronic transport and that the Landauer-Büttiker term becomes dominant whenever the incoherent term proportional to $\Gamma_{ee}$ cancels. We are thus able to capture the essential feature of the experimental result and to describe it, namely, the behavior of a single electron transistor with Coloumb diamonds associated with transitions between singly and doubly occupied states.
Finally, while our calculations were performed for a specific arrangement of nanoribbon width and length, we expect this behavior to be quite general. In fact, an increase in ribbon width - within the same ribbon family - would lead to a larger number of states closer to the Fermi level, but a stronger NT/NR coupling. This leads to the presence of a larger number of Coulomb diamonds within the same window of applied gate potential. At the same time, the stronger coupling would eventually lead back to a single particle description. This interplay could pave the way for engineering devices with appropriate charging energies for applications.

\section{Data availability statement}

The data that support the findings of this study are available from the author, W. F. dos Santos, upon reasonable request.

\section{Declaration of conflict of interest statement}

The authors wish to confirm that there are no known conflicts of interest associated with this publication and there has been no significant financial support for this work that could have influenced its outcome.

\section{Acknowledgements}
The authors acknowledge financial support by CAPES Brazil, CNPq and FAPESP (grant numbers
2023/11751-9,2023/09820-2,2021/14335-0 and 2017/02317-2). This work used the computational
resources from the \textit{Centro Nacional de Processamento de Alto Desempenho
em São Paulo} (CENAPAD-SP) and GRID-UNESP.

\section{References}
\bibliography{iopart-num}

\end{document}